\documentclass{emulateapj}
\usepackage{graphicx}
\usepackage{epstopdf} 
\usepackage{amsmath}

\usepackage{natbib, natbibspacing}

\begin{document}

\title{Adaptive Optics Images of Kepler Objects of Interest\altaffilmark{1}}

\author{E. R. Adams\altaffilmark{2},  D. R. Ciardi\altaffilmark{3}, A. K. Dupree\altaffilmark{2},  T. N. Gautier III\altaffilmark{4}, C. Kulesa\altaffilmark{5}, D. McCarthy\altaffilmark{5}}

\altaffiltext{1}{Based on observations obtained at the MMT Observatory, a joint facility of the Smithsonian Institution and the University of Arizona.}
\altaffiltext{2}{Harvard-Smithsonian Center for Astrophysics, 60 Garden St., Cambridge, MA, 02138.}
\altaffiltext{3}{NASA Exoplanet Science Institute, 770 South Wilson Avenue, Pasadena, CA, 91125}
\altaffiltext{4}{Jet Propulsion Laboratory, California Institute of Technology, 4800 Oak Grove Dr, Pasadena, CA 91109}
\altaffiltext{5}{Steward Observatory, The University of Arizona, 933 N. Cherry Ave, Tucson, AZ, 85721.}

\begin{abstract}

All transiting planets are at risk of contamination by blends with nearby, unresolved stars. Blends dilute the transit signal, causing the planet to appear smaller than it really is, or produce a false positive detection when the target star is blended with eclipsing binary stars. This paper reports on high spatial-resolution adaptive optics images of 90 \emph{Kepler} planetary candidates. Companion stars are detected as close as 0.1\arcsec\ from the target star. Images were taken in the near-infrared ($J$ and $Ks$ bands) with ARIES on the MMT and PHARO on the Palomar Hale 200-inch. Most objects (60\%) have at least one star within 6\arcsec\ separation and a magnitude difference of 9. Eighteen objects (20\%) have at least one companion within 2\arcsec\ of the target star; 6 companions (7\%) are closer than 0.5\arcsec. Most of these companions were previously unknown, and the associated planetary candidates should receive additional scrutiny. Limits are placed on the presence of additional companions for every system observed, which can be used to validate planets statistically using the BLENDER method. Validation is particularly critical for low-mass, potentially Earth-like worlds, which are not detectable with current-generation radial velocity techniques. High-resolution images are thus a crucial component of any transit follow-up program.

\end{abstract}


\section{Introduction}

Two methods have been responsible for the vast majority of extrasolar planet discoveries. Radial velocity observations are used to determine the planetary mass from the amplitude of the shift in spectral lines due to the planet's gravitational pull on the star, while transit photometry gives the planetary radius based on the amount of light blocked as the planet transits across the host star. Both of these methods, however, are vulnerable to unresolved light from nearby stars, whether the contaminating star is a bound companion or a chance alignment. When close companions fall within the same spectral slit or photometric aperture as the target, the resulting blend distorts the derived planetary parameters and sometimes creates false positive signals. 

Accounting for nearby stars is particularly important for transiting planets that lack corresponding radial velocity measurements and hence have no confirmation of planetary mass. This paper focuses on the transiting planet candidates identified by the \emph{Kepler} space mission, currently numbering over 2300 \citep{Batalha2012}.  Many of these objects do not have mass estimates from radial velocity measurements because of the amount of observing time required, particularly for small planets around relatively faint stars. Current radial velocity instruments cannot detect planets around the size of the Earth, and thus are not useful for confirming objects of this size that have recently been found with transit photometry \citep[e.g.][]{Fressin2012a}. Another method is needed to confirm these types of planets.

The best way to account for nearby stars is provided by high-resolution imaging. A variety of methods have been used to examine planet-hosting stars, such as lucky imaging \citep[e.g.][]{Daemgen2009}, speckle imaging \citep[e.g][]{Howell2011}, and adaptive optics (this work). High-resolution images are used to (1) rule out false positive scenarios caused by background blends \citep[e.g.][]{Hartman2011a}, (2) estimate the dilution of the transit light curve caused by additional faint stars in the aperture \citep[e.g.][]{Buchhave2011,Demory2011}, and (3) confirm statistically that a candidate is probably a planet even without radial velocity measurements, by calculating the likelihood of all false positive scenarios using the BLENDER method \citep[e.g.][]{Torres2011, Fressin2011, Ballard2011, Fressin2012a}.

 \subsection{False positives}

For transiting planet detections, the main source of false positives is a blend of the target star with an eclipsing binary star. The \emph{Kepler} team searches for many signatures of such blends, include examining the light curves for V-shapes, discrepancies between odd and even transits, and signs of secondary occultations that are too deep to be planetary \citep{Batalha2010}. In addition, the center of light for each target is tracked in and out of transit. If the centroid shifts significantly, this indicates that the source of the transit signal is around a different star. At a minimum, the derived planetary parameters must be completely reevaluated, sometimes resulting in a false positive \citep{Jenkins2010}. This extensive vetting effort is partly why the candidates announced by the \emph{Kepler} team are thought to have a low false positive rate, variously estimated from $5-20\%$ \citep{Borucki2011, Morton2011, Wolfgang2011}. Finding which of the announced candidates are actually false positives requires extensive follow-up efforts, including ground- and space-based high-resolution imaging.

For the purposes of this paper, a false positive is defined as a transit signal that is not produced by a planetary-mass object around the proposed target star. Some false positives may actually be larger planets around a fainter star that is blended with the brighter target star.

\subsection{Dilution corrections}

Additional stars in the photometric aperture will dilute the transit signal and distort the measured planetary radius. The amount of dilution depends on the brightness of the background star and how much of its light falls in the aperture. A \emph{Kepler} pixel is about 4\arcsec\ on a side, while a typical \emph{Kepler} aperture is 12\arcsec\ across, and existing catalogs typically do not contain companions more than a few magnitudes fainter or closer than a few arcseconds. Thus, it is critical to probe as close to the star as possible to find nearby companions that might challenge the identity and parameters of the star hosting the transit signature. 

It is particularly useful to obtain dilution corrections in more than one wavelength. An important component of the \emph{Kepler} follow-up is transit observations with the near-infrared \emph{Spitzer} satellite. Most false-positive blend scenarios produce a color-dependent transit-like signature. Thus, an object is more likely to be a planet if it has the same depth at both the \emph{Kepler} bandpass and, for instance, at 4.5 $\mu m$ with \emph{Spitzer}. If the depth of the transit does vary with wavelength, however, it is vital to have resolved images of nearby stellar companions to test whether the depth variation is more consistent with a false positive or with a genuine planetary transit that is being diluted by the companion star(s). 

\subsection{Planetary validation}

If no companions are detected with direct imaging, we can place strict limits on any remaining companions using BLENDER. The BLENDER method has been developed to combine the constraints on false positive scenarios placed by all available radial velocity, photometric, and imaging measurements \citep{Torres2011}. Strong limits on the allowed magnitude difference of undetected stars within a few tenths of an arcsecond are particularly useful. The probability of a false positive scenario scales directly with the area in which a background blend can exist within the limits placed by AO imaging. If the combined probability of any allowed background blend is much less than the planet occurence frequency, the candidate is said to be statistically validated. 

Validation is particularly critical for the smallest objects, which cannot be confirmed using radial velocity techniques. Several planets from 1-2 $R_E$ have been statistically validated using the BLENDER method, include Kepler-9d \citep{Torres2011}, Kepler-10c \citep{Fressin2011}, Kepler-19b \citep{Ballard2011}, Kepler-22b \citep{Borucki2012}, and Kepler-20e and f \citep{Fressin2012a}.

This paper reports on observations of 90 candidate planetary objects, or Kepler Objects of Interest (KOIs). The observations are described in Section~\ref{section:obs}. The limits placed on observable companions are reported in Section~\ref{section:limits}. A list of all companion stars within 6\arcsec\ of the targets is reported in  Section~\ref{section:freq}, along with a discussion of the frequency and implications of the companion stars.

\section{Observations and data analysis}
\label{section:obs}

All observations were made during the 2009 and 2010 seasons (roughly May to November). Data were obtained over four nights on the MMT and seven nights on Palomar, as listed in Table~\ref{table:obs}, with observations of 90 unique KOIs. 

When possible, objects were observed in both $J$ and $Ks$ bands; many of these objects have also been observed using speckle imaging in optical wavelengths \citep{Howell2011}. Observations in multiple wavelengths are particularly useful for estimating the spectral type of any detected companions, and for converting the observed delta-magnitude into \emph{Kepler's} broad optical band  ($Kp$). All observations in the this paper are relative photometry, so the absolute $J$ and $Ks$ magnitudes are found by adding the 2MASS catalog magnitude to the relative magnitudes observed by ARIES and PHARO. The resulting absolute magnitudes are converted to $Kp$-band magnitudes using the fifth-order polynomial fits for dwarf stars presented in Appendix A of \citet{Howell2012}:

\begin{equation}
Kp-J =\begin{cases}
\begin{split}
-398.04666+149.08127J\\
- 21.952130 J^2+1.5968619 J^3\\
 - 0.057478947 J^4+0.00082033223J^5,
\end{split}
& J \le 16.7, \\
0.1918 + 0.08156J, 			& J > 16.7
\end{cases}
\label{eqnJ}
\end{equation}

and

\begin{equation}
Kp-Ks =\begin{cases}
\begin{split}
-643.05169+246.00603\emph{Ks}\\
-37.136501\emph{Ks}^2 +2.7802622\emph{Ks}^3\\
 - 0.10349091{Ks}^4 +0.0015364343\emph{Ks}^5,
\end{split}
& Ks \le 15.4, \\
-2.7284 + 0.3311\emph{Ks}, 			& Ks > 15.4.
\end{cases}
\label{eqnKs}
\end{equation}

The single-filter conversions in Equations~\ref{eqnJ} and \ref{eqnKs} yield $Kp$ magnitude estimates that are accurate to approximately $0.6$-$0.8$ mag, and are used for all magnitude limit calculations. For objects detected in both $J$ and $Ks$, a better estimate of the $Kp$ magnitude can be obtained by using the dual-filter conversions, which yield $Kp$ magnitudes accurate to about $0.05$ mag \citep{Howell2012}.

\begin{equation}
Kp - K s = \begin{cases}
\begin{split}
0.314377 + 3.85667x\\
+ 3.176111x^2 - 25.3126x^3 \\
+ 40.7221x^4 - 19.2112x^5,
\end{split}
& \mbox{Dwarfs}, \\

0.42443603 + 3.7937617x\\
 - 2.3267277x^2 + 1.4602553x^3					&\mbox{Giants}
\end{cases}
\label{eqnBoth}
\end{equation}
where $x = J - \emph{Ks}$.

\subsection{ARIES}

Observations with ARIES were taken on four nights between Nov 2009 and Sep 2010. The Arizona Infrared imager and Echelle Spectrograph (ARIES) on the 6.5m MMT telescope can provide diffraction-limited imaging in the $J$ and $Ks$ bands \citep{McCarthy1998}. ARIES is fed by the adaptive secondary AO system. Most KOIs were imaged in the f/30 mode, with a field of view of 20\arcsec\ $\times$ 20\arcsec\ and a resolution of 0.02085\arcsec\ per pixel. In poor seeing, the f/15 mode is used, with a field of view of 40\arcsec\ $\times$ 40\arcsec\ and a resolution of 0.0417\arcsec\ per pixel. The adaptive optics system in all cases guided on the primary target. The median width of the central cores of the point spread functions were $0.25\arcsec$ at $J$ and $0.14\arcsec$ at $Ks$, with a best resolution of 0.1\arcsec\ at J and 0.09\arcsec\ at Ks. Under good conditions in May 2010 (uncorrected seeing of 0.5\arcsec\ at Ks), the Strehl ratios were measured at 0.3 in Ks and 0.05 in J.

For each KOI, at least one set of 16 images on a 4-point dither pattern were observed in both $J$ and $Ks$. In May and Sep 2010, a random jitter was added to the dither position, which had steps of 2 or 4\arcsec\ . Integration times varied from 0.9 to 30 s depending on stellar magnitude; in some cases, more than one set of 16 images were taken. The images for each filter were calibrated using standard IRAF procedures\footnote{http://iraf.noao.edu/}, and combined and sky-subtracted using the \emph{xmosaic} function in the \emph{xdimsum} package. The images taken in 2009-2010 have a slight differential rotation in them, which was too small to require correction near the target, but causes stars near the edges of the field to be smeared out when stacked. The orientations of the fields are estimated from the dither pattern, and are only accurate to within a few degrees.

\subsection{PHARO}

Near-infrared adaptive optics imaging was obtained on the nights of
07-10 Sep 2009 and 01-03 July 2010 UT with the Palomar Hale 200-inch
telescope and the PHARO near-infrared camera \citep{Hayward2001} behind
the Palomar adaptive optics system \citep{Troy2000}.  PHARO, a
$1024\times20124$ HgCdTe infrared array, was used in 0.0251\arcsec\  pixel$^{-1}$
mode yielding a field of view of $25\arcsec$.  The KOIs observed
in 2009 were imaged only in the $J$ filter ($\lambda_0 = 1.25~\mu$m)
while the KOIs observed in 2010 were imaged in the in both the $J$ and
$K_{s}$ ($\lambda_0 = 2.145~\mu$m) filters. All the data were
collected in a standard 5-point quincunx dither pattern of 5\arcsec\
steps interlaced with an off-source ($60\arcsec$) sky dither
pattern. Individual integrations times varied depending on the brightness of
the KOIs, from 1.4 to 69 s, and were aimed at detecting sources 9 magnitudes fainter than
the target in $J$ and 8 magnitudes fainter in $Ks$ ($5\sigma$).  The
individual frames were reduced with a custom set of IDL routines written
for the PHARO camera and were combined into a single final image.  In
all cases, the adaptive optics system guided on the primary target
itself. The median width of the central cores of the point spread functions were 0.08\arcsec\ at $J$ and 0.1\arcsec at $Ks$, with a best resolution at 0.05\arcsec\ at J and 0.09\arcsec\ at Ks. The Strehl ratio for good images is 0.1-0.15 in $J$ and 0.35-0.5 in $Ks$.

\section{Detection limits}
\label{section:limits}

All objects were identified by manual inspection, which was more efficient at weeding out spurious signals and artifacts than automatic detection methods. The magnitude of a companion was estimated using the IRAF routine \emph{phot} using a 5 pixel aperture (large enough to capture most of the point spread function, or PSF, without also including light from all but the closest companions). In a few cases, PSF fitting was used on very close companions (such as K00098, separation=0.3\arcsec). 

Limits on undetected stars are estimated as follows. A series of concentric annuli are drawn around the star, and the standard deviation of the background counts is calculated for each annulus. A star is considered detectable if its peak signal is more than 5 times the standard deviation above the background level. The magnitude of this star is reported as the detection limit at the distance of the center of the annulus. Limits are reported for distances from  0.1-4\arcsec\ in Table~\ref{table:limits}. The 4\arcsec\ level can also be applied toward more distant objects.

\begin{figure}
\includegraphics*[scale=0.5]{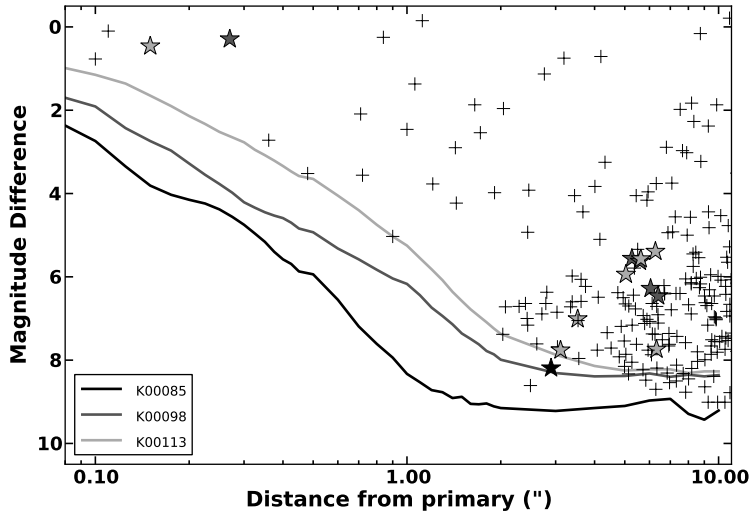}
\caption{Detected companions and limits on additional stars in $Ks$-band. All stars detected near 90 targets are shown as plus signs or stars. Detection limits and all known companions are shown for three systems: K00085 (black), one companion at 2\arcsec; K00098 (dark gray), one close companion (0.3\arcsec) and several more distant; and K00113 (light gray), one close companion (0.15\arcsec) and several more distant. Note that the detection limits vary from system to system by several magnitudes depending the total integration time and the observational conditions.}
\label{fig:complimits}
\end{figure}

The innermost detectable object is a function of the observed PSF of the target star. The best FWHM achieved for targets in $J$ band was 0.1\arcsec\ for ARIES and 0.05\arcsec\ for PHARO, while both instruments reached 0.09\arcsec\ in $Ks$. However, poor weather and problems with the AO systems often caused excursions well above that level. The magnitude limits for each KOI are shown in Table~\ref{table:limits}. The limits on a few examples are shown in Figure~\ref{fig:complimits}, along with a scatter plot of the companions detected near all targets.

\section{Frequency and implications of companions}
\label{section:freq}

Additional faint stars are common near \emph{Kepler} targets. Over half (53/90, or nearly 60\%) of the targets imaged have at least one companion within 6\arcsec. All of the stars with companions are listed in Table~\ref{table:stats}, while a list of the relative magnitudes, distances and position angles is in Table~\ref{table:compstars}. Many of these objects are very faint (down to 10 magnitudes fainter in $Ks$, and typically even fainter visible magnitudes), and so have little dilution effect on the \emph{Kepler} light curves. Being able to say for certain that there are no brighter objects present lends confidence to the stated planetary parameters.

\begin{figure}
\includegraphics*[scale=0.3]{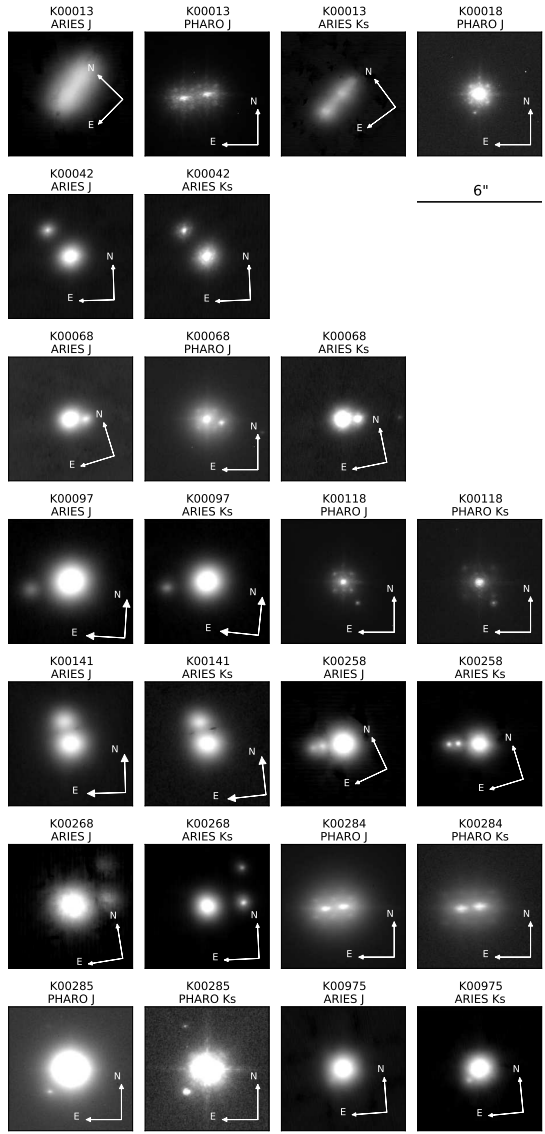}
\caption{Kepler Objects of Interest (KOIs) with at least one companion between 0.5 and 2\arcsec. Each field of view is 6\arcsec, 1.5 times the width of a Kepler pixel. The color scale for all images is logarithmic and chosen to highlight the companion star(s). Four objects have two companions within 3\arcsec: K00068 (the fainter is near the north arrow); K00258; K00268; and K00285. Note the artifacts  in K00018 (real companion is due south) and K00118 (companion to SW). The blurred ARIES images of K00013 are due to the AO system having trouble locking on one of the similarly bright stars. The companion to K00975 is very red, and only marginally detected at $J$.}
\label{fig:comps}
\end{figure}

\begin{figure}
\includegraphics*[scale=0.3]{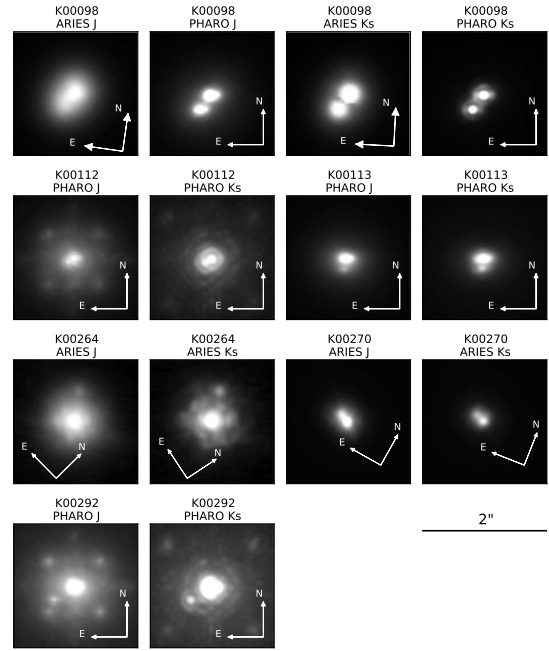}
\caption{Kepler Objects of Interest (KOIs) with companions closer than 0.5\arcsec. Each field of view is 2\arcsec, half the width of a Kepler pixel. The color scale is logarithmic for K00112, K00264, and K00292, to highlight the core of the PHARO and ARIES PSFs, and is linear for K00098, K00113, and K00270 in order to make the companion star visible. K00264 (ARIES, $Ks$) shows the blobs and speckles that come with imperfect AO correction. The real companion is identifiable because it lies just outside the core of the PSF. K00292 (PHARO, $J$) demonstrates the quincunx pattern of the PHARO PSF, with the real companion (to the southeast of the target) identifiable because it only appears on images of this star and not of other targets. }
\label{fig:compsZoom}
\end{figure}

Close companions, within 2\arcsec, are of particular concern, since they are within the same \emph{Kepler} pixel as the target, and may not produce a detectable centroid shift. Of the objects presented here, 20\% of objects imaged have at least one companion within 2\arcsec, and 7\% have one within 0.5\arcsec. The images of twelve KOIs with detected companions between 0.5-2\arcsec\ are shown in Figure~\ref{fig:comps}, while six objects with companions closer than 0.5\arcsec\ are shown in Figure~\ref{fig:compsZoom}.  

This subset of planetary candidates with close stellar companions should be carefully examined for false positives; the list in Table~\ref{table:compstars} in fact contains several objects that have since been identified as likely false positives by the \emph{Kepler} team, including K00068, K00076, K00088, K00264, and K00266. 

On the other hand, several transiting candidates around stars with close stellar companions have been confirmed to be planetary. Knowledge of the additional star has been incorporated into the derivation of the correct planetary parameters. Several examples of confirmed planetary systems are K00098 \citep[aka Kepler-14,][]{Buchhave2011}, K00097 \citep[aka Kepler-7,][]{Demory2011}, and K00975 \citep[aka Kepler-21,][]{Howell2012}. For some objects, such as K00097, the corrections were relatively minor, at the level of a few percent. However, for K00098, the companion was within 0.3\arcsec\ and only a few tenths of a magnitude fainter, and the dilution corrections were substantial: the radius increased by 10\%, the mass by 60\% (since the stellar spectra were also blended), and the density changed by 25\% \citep{Buchhave2011}. Without high resolution images, we would have had a very inaccurate picture of this planet.

Not surprisingly, given the generally red colors of faint objects, more apparent binaries occur in the infrared than in the visible. High-resolution optical speckle images al by \citet{Howell2011} found that only 10/156 stars, or 6.4\%, had companions within 2\arcsec\ and down to 4 magnitudes fainter.  (No attempts have been made to correct for the selection biases in the objects that were selected for follow-up observations.)

It is unknown which, if any, of the detected companions are bound to their targets. At present, we lack the time baseline needed to detect proper motion for any of the closest companions. Two statistical arguments can be made to argue that many of the closest stars are likely to be bound. The first is to note that if all of the companions detected were unconnected background or foreground objects, then we would expect to find 9 times as many objects within 6\arcsec\ as within 2\arcsec, However, we actually find only 3 times as many objects (53 vs. 18), and the ratio would be smaller if it included the the closest objects that are missed because they are within the stellar PSF.

\begin{figure}
\includegraphics*[scale=0.5]{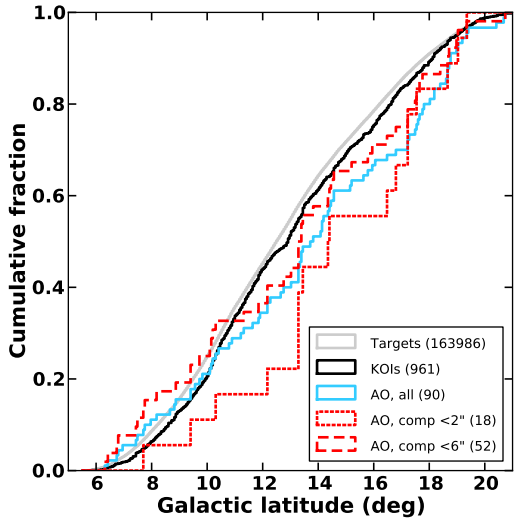}
\caption{Cumulative distribution of Kepler objects vs. Galactic latitude. All stars targeted for transit searches are shown in light gray, while the set of KOIs from \citet{Borucki2011} is in black; the numbers in parentheses are the number of objects in each group. The close match between targets and KOIs means the full KOI list is evenly distributed with latitude. The 90 objects targeted for AO followup (thick blue in color version or dark grey in black and white) are not an unbiased subset, and happen to have a deficit of objects at intermediate latitudes. Also shown are two subsets of AO objects, those with detected companions (of any magnitude) within 6\arcsec\ (dashed red line) and within 2\arcsec\ (dotted red line). A slight bias toward lower Galactic latitude is seen in the stars with more distant (6\arcsec) companions, while the opposite bias is hinted at for closer (2\arcsec) companions. Though the number of objects are still low, this may indicate that the closest objects are not background companions, and are more likely to be physically bound.}
\label{fig:lat}
\end{figure}

The second statistical argument is to compare the Galactic latitudes of KOIs with detected companions within 2 and 6\arcsec, as shown in Figure~\ref{fig:lat}. The AO targets with companions within 6\arcsec\ are somewhat more likely to appear at low Galactic latitudes, indicating that some of them are likely background blends, but no such correlation can be seen with the (admittedly small) sample of close companions (within 2\arcsec). Thus many of the closest objects may be part of physical binary systems, but further observations are required to determine which ones they may be.

\section{Conclusions}
\label{section:ogle56conclusions}

High-resolution, adaptive optics images of 90 \emph{Kepler} planetary candidates have been obtained. A list of all companions within 6\arcsec\ is provided, with measured magnitudes in $J$ and/or $Ks$ band and calculated \emph{Kepler} magnitudes. Limits on additional companions from 0.1 to 4\arcsec\ are also given for each target, and can be used to calculate the probability of a remaining undetected blend.

Roughly 20\% of the objects imaged have at least one companion within 2\arcsec, and 7\% have one within 0.5\arcsec. Over half have a more distant companion at a distance of 6\arcsec. Although small number statistics apply, the objects within 2\arcsec\ appear uncorrelated with Galactic latitude, making it more likely that they represent physically bound (though still distant) companion stars.  

Even if false positive blends can be ruled out, corrections to the planetary parameters based on nearby stars can range from a few to tens of percents, making high resolution images an important tool to understanding the true sizes of other discovered worlds.

\acknowledgements




\newpage

\begin{deluxetable*}{llllllllllll}
\tablewidth{0pt}
\tabletypesize{\scriptsize}
\tablecaption{Summary of observations}
\tablehead{ Date (UTC) 	& Instrument	& Number of KOI\tablenotemark{a}	& Notes}
\startdata
2009 Nov 11					& ARIES		& 4 (J, Ks) + 4 (Ks)			& f/15 mode (unstable seeing)		\\
2010 May 2 					& ARIES		& 6 (J, Ks)	\tablenotemark{b}	& f/30 mode; electronics problems	\\
2010 May 3 					& ARIES		& 17 (J, Ks)				& f/30 mode; approached diffraction limit		\\
2010 Sep 24					& ARIES		& 3 (J, Ks)					& f/30 mode		\\
2010 Sep 26					& ARIES		& 9 (J, Ks)					& f/30 mode	\\
2009 Sep 07					&  PHARO	& 6 (J)					& 1.5\arcsec\ uncorrected seeing at J \\
2009 Sep 08					&  PHARO	& 12 (J)					& 2.0\arcsec\ uncorrected seeing at J \\
2009 Sep 09					&  PHARO	& 10 (J)					& 0.8\arcsec\ uncorrected seeing at J \\
2009 Sep 10					&  PHARO	& 7 (J)					& 0.8\arcsec\ uncorrected seeing at J \\
2010 Ju1 01					&  PHARO	& 11 (J, Ks) + 1 (Ks)			& 2.0\arcsec\ uncorrected seeing at J \\
2010 Jul 02					&  PHARO	& 11 (J, Ks)				& 1.3\arcsec\ uncorrected seeing at J\\
2010 Jul 03					&  PHARO	& 8 (J)					& 1.5\arcsec\ uncorrected seeing at J 
\enddata     	
\tablenotetext{a} {Some objects were observed by both instruments and/or on more than one night, so numbers do not add up to 90 objects.}
\tablenotetext{b} {Problems with instrument; all objects re-observed on May 3.}
\label{table:obs}
\end{deluxetable*}

\begin{deluxetable*}{c c c c   c c c c c c }
\tablecolumns{11} 
\tablewidth{0pt}
\tabletypesize{\scriptsize}
\tablecaption{Limits on nearby stars for all KOIs}
\tablehead{ 
KOI   & Instr. 	& Filter 	& FWHM 		&  \multicolumn{6}{c}{Limiting $\Delta$ Mag for annulus centered at} \\ 
		  &   		& 		& (\arcsec)	& 0.1\arcsec& 0.2\arcsec &  0.5\arcsec &  1\arcsec & 2\arcsec &  4\arcsec  
}
\startdata
K00005	&PHARO	&J	&0.23	&--	&--	&3.1	&5.1	&7.3	&8.6 \\
	&	&Kep	&	&--	&--	&3.3	&5.4	&7.9	&9.2 \\
K00007	&PHARO	&J	&0.05	&5.0	&5.7	&6.6	&8.7	&9.7	&9.9 \\
	&	&Kep	&	&5.4	&6.1	&7.2	&9.4	&10.5	&10.8 \\
K00008	&PHARO	&J	&0.08	&3.4	&4.2	&5.1	&7.3	&9.2	&9.7 \\
	&	&Kep	&	&3.7	&4.6	&5.6	&8.0	&10.0	&10.6 \\
K00010	&PHARO	&J	&0.08	&2.9	&3.7	&4.6	&6.4	&8.5	&9.4 \\
	&	&Kep	&	&3.4	&4.2	&5.2	&7.1	&9.4	&10.4 \\
K00011	&PHARO	&J	&0.06	&4.3	&5.2	&6.4	&8.4	&9.3	&9.5 \\
	&	&Kep	&	&4.5	&5.5	&6.8	&9.0	&10.0	&10.2 \\
K00013	&PHARO	&J	&0.14	&--	&3.7	&4.3	&6.4	&8.3	&9.4 \\
	&	&Kep	&	&--	&4.5	&5.1	&7.4	&9.5	&10.7 \\
K00013	&ARIES	&J	&1.24	&--	&--	&--	&--	&4.4	&7.0 \\
	&	&Kep	&	&--	&--	&--	&--	&5.3	&8.0 \\
K00013	&ARIES	&Ks	&1.27	&--	&--	&--	&--	&5.2	&5.9 \\
	&	&Kep	&	&--	&--	&--	&--	&6.7	&7.7 \\
K00017	&PHARO	&J	&0.09	&2.4	&3.1	&4.4	&6.3	&8.4	&9.0 \\
	&	&Kep	&	&2.4	&3.3	&4.6	&6.7	&8.9	&9.6 \\
K00018	&PHARO	&J	&0.06	&3.7	&4.5	&5.5	&7.7	&9.2	&9.5 \\
	&	&Kep	&	&3.9	&4.8	&5.9	&8.3	&9.9	&10.2 \\
K00020	&PHARO	&J	&0.1	&2.6	&3.2	&4.3	&6.3	&8.5	&9.4 \\
	&	&Kep	&	&3.0	&3.6	&4.8	&7.0	&9.4	&10.3 
\enddata     
\tablenotetext{a} {Full table available as online supplement.}
\label{table:limits}
\end{deluxetable*}

\begin{deluxetable*}{p{3cm} p{1cm} p{8cm} }
\tablewidth{0pt}
\tabletypesize{\scriptsize}
\tablecaption{Companion statistics}
\tablehead{Number of companions	& 	Targets\tablenotemark{a} & KOI}
\startdata
None within 6\arcsec		&	37		& 5, 7, 11, 22, 28, 41, 64, 69, 72, 84, 103, 109, 111, 127, 180, 92, 104, 117, 244, 247, 265, 271, 275, 281, 313, 365, 116, 245, 246, 257, 260, 262, 274, 76, 82, 94, 974 \\
1+  within 6\arcsec			& 	53		& 8, 10, 13, 17, 18, 20, 42, 68, 70, 75, 85, 87, 88, 97, 98, 102, 106, 108, 112, 113, 115, 118, 121, 122, 123, 124, 126, 137, 141, 148, 153, 249, 251, 258, 261, 263, 264, 266, 268, 269, 270, 273, 283, 284, 285, 292, 303, 306, 316, 364, 372, 377, 975\\
2+ within 6\arcsec			&	25		& 8, 10, 18, 68, 87, 98, 102, 106, 108, 113, 115, 121, 123, 126, 137, 148, 251, 258, 261, 268, 285, 306, 364, 372, 377 \\
3+  within 6\arcsec			&	10		& 18, 68, 106, 113, 126, 137, 148, 306, 364, 372 \\
\hline
1+ within 2\arcsec			& 	18		& 13, 18, 42, 68, 97, 98, 112, 113, 118, 141, 258, 264, 268, 270, 284, 285, 292, 975 \\
2 within 2\arcsec			& 	1		& 258\\
\hline
1 within 0.5\arcsec			& 	6		& 98, 112, 113, 264, 270, 292 
\enddata     	
\tablenotetext{a} {Number of Kepler Objects of Interest (KOIs) with companions at a given distance (90 total targets).}
\label{table:stats}
\end{deluxetable*}

\newpage

\begin{deluxetable*}{crrrrcc  |  rrr | rrr   |  c}
\tablewidth{0pt}
\tablecaption{Stars within 6\arcsec\ of \emph{Kepler} planetary candidates}
\tablehead{
	&		&	&2MASS	&2MASS	&	&  & \multicolumn{3}{c}{J} 	& \multicolumn{3}{c}{Ks}  & Est Kep\\
KOI	&KeplerID	 &Kep	& J	& K	&Inst.\tablenotemark{a}	& Star\tablenotemark{b} &Dist(\arcsec)	&PA($^{\circ}$)\tablenotemark{c}	&$\Delta$ Mag\tablenotemark{d} &Dist(\arcsec)	&PA($^{\circ}$)\tablenotemark{c}	 &$\Delta$ Mag\tablenotemark{d} &  Mag\tablenotemark{e}
		}
\startdata
K00008	&5903312	&12.450	&11.371	&11.039	&P	&1	&5.15	&132.0	&6.7	&	&	&	&19.7	\\
	&	&	&	&	&P	&2	&5.84	&173.6	&8.5	&	&	&	&21.7	\\
\hline
K00010	&6922244	&13.563	&12.576	&12.292	&P	&1	&3.04	&94.3	&7.7	&	&	&	&22.1	\\
	&	&	&	&	&P	&2	&3.74	&89.3	&6.2	&	&	&	&20.5	\\
\hline
K00013	&9941662	&9.958	&9.465	&9.425	&A	&1	&0.8\tablenotemark{f}	&78.5	&0.1	&1.12	&102.4	&-0.0	&10.3	\\
	&	&	&	&	&P	&1	&1.12	&99.4	&-0.2	&	&	&	&10.5	\\
\hline
K00017	&10874614	&13.303	&12.001	&11.634	&P	&1	&4.01	&39.9	&3.8	&	&	&	&17.3	\\
\hline
K00018	&8191672	&13.369	&12.115	&11.769	&P	&1	&0.9	&166.9	&5.0	&	&	&	&18.7	\\
	&	&	&	&	&P	&2	&3.39	&110.1	&6.0	&	&	&	&19.8	\\
	&	&	&	&	&P	&3	&4.94	&149.0	&6.5	&	&	&	&20.3	\\
\hline
K00020	&11804465	&13.438	&12.406	&12.066	&P	&1	&5.04	&139.6	&7.9	&	&	&	&22.2	\\
\hline
K00042	&8866102	&9.364	&8.416	&8.140	&A	&1	&1.63	&39.4	&2.2	&1.65	&39.0	&1.9	&12.1	\\
\hline
K00068	&8669092	&12.733	&11.740	&11.401	&A	&1	&0.71	&107.6	&2.0	&0.72	&100.2	&1.8	&15.1	\\
	&	&	&	&	&A	&2	&2.73	&107.6	&7.2	&2.71	&99.8	&6.2	&16.2	\\
	&	&	&	&	&A	&3	&3.41	&12.6	&6.4	&3.38	&4.5	&5.8	&20.6	\\
	&	&	&	&	&P	&1	&0.71	&103.6	&2.1	&	&	&	&15.2	\\
	&	&	&	&	&P	&2	&2.71	&103.5	&6.9	&	&	&	&20.4	\\
	&	&	&	&	&P	&3	&3.35	&8.6	&6.7	&	&	&	&20.1	\\
\hline
K00070	&6850504	&12.498	&11.252	&10.871	&P	&1	&3.67	&51.9	&4.4	&	&	&	&17.1	\\
\hline
K00075	&7199397	&10.775	&9.692	&9.387	&A	&1	&3.42	&127.6	&6.7	&3.43	&123.5	&6.5	&17.7	\\
	&	&	&	&	&P	&1	&3.39	&125.0	&6.6	&	&	&	&17.8	\\
\hline
K00085	&5866724	&11.018	&10.066	&9.805	&A	&1	&	&	&	&2.9	&56.1	&8.2	&21.2	\\
\hline
K00087	&10593626	&11.664	&10.522	&10.151	&P	&1	&5.37	&177.4	&5.9	&5.37	&177.4	&5.8	&17.7	\\
	&	&	&	&	&P	&2	&5.4	&75.2	&7.3	&5.39	&75.1	&6.8	&20.1	\\
\hline
K00097	&5780885	&12.885	&11.833	&11.535	&A	&1	&1.9	&105.1	&4.0	&1.91	&105.1	&4.0	&16.9	\\
\hline
K00098	&10264660	&12.128	&11.201	&10.987	&A	&1	&0.26\tablenotemark{g}	&136.6	&0.40\tablenotemark{g}	&0.27\tablenotemark{g}	&146.6	&0.49\tablenotemark{g}	&12.0	\\
	&	&	&	&	&A	&2	&5.6	&49.3	&6.3	&5.59	&50.0	&5.6	&19.9	\\
	&	&	&	&	&P	&1	&0.27	&143.7	&0.3	&0.28	&143.5	&0.4	&12.2	\\
	&	&	&	&	&P	&2	&5.27	&54.1	&6.2	&5.27	&54.1	&5.6	&19.5	\\
\hline
K00102	&8456679	&12.566	&11.397	&11.054	&P	&1	&2.76	&137.8	&1.1	&	&	&	&13.9	\\
	&	&	&	&	&P	&2	&5.45	&133.2	&7.6	&	&	&	&20.7	\\
\hline
K00106	&10489525	&12.775	&11.775	&11.516	&P	&1	&2.07	&109.3	&6.7	&	&	&	&20.2	\\
	&	&	&	&	&P	&2	&5.28	&142.5	&7.2	&	&	&	&20.7	\\
	&	&	&	&	&P	&3	&5.53	&129.5	&7.1	&	&	&	&20.6	\\
\hline
K00108	&4914423	&12.287	&11.192	&10.872	&P	&1	&2.44	&74.9	&7.2	&	&	&	&20.1	\\
	&	&	&	&	&P	&2	&4.87	&112.4	&7.2	&	&	&	&20.1	\\
\hline
K00112	&10984090	&12.772	&11.698	&11.367	&P	&1	&0.1	&116.9	&0.8	&0.11	&119.7	&0.8	&13.6	\\
\hline
K00113	&2306756	&12.394	&11.150	&10.720	&P	&1	&0.17	&165.5	&0.6	&0.15	&167.3	&0.5	&13.1	\\
	&	&	&	&	&P	&2	&3.11	&68.1	&7.8	&3.12	&68.1	&7.8	&20.2	\\
	&	&	&	&	&P	&3	&3.53	&46.7	&7.4	&3.53	&46.6	&7.0	&20.7	\\
	&	&	&	&	&P	&4	&5.05	&96.5	&6.0	&5.04	&96.5	&5.9	&18.5	\\
	&	&	&	&	&P	&5	&5.63	&176.8	&6.1	&5.63	&176.7	&5.6	&19.7	\\
\hline
K00115	&9579641	&12.791	&11.810	&11.503	&A	&1	&	&	&	&2.43	&139.2	&7.0	&21.9	\\
	&	&	&	&	&A	&2	&	&	&	&4.16	&162.0	&5.1	&19.4	\\
\hline
K00118	&3531558	&12.377	&11.273	&10.897	&P	&1	&1.21	&145.8	&4.0	&1.21	&145.8	&3.8	&16.7	\\
\hline
K00121	&3247396	&12.759	&11.723	&11.436	&P	&1	&2.81	&168.4	&6.4	&	&	&	&19.8	\\
	&	&	&	&	&P	&2	&3.62	&171.2	&6.6	&	&	&	&20.0	\\
\hline
K00122	&8349582	&12.346	&11.210	&10.801	&P	&1	&4.11	&148.8	&6.5	&	&	&	&19.3	\\
\hline
K00123	&5094751	&12.365	&11.314	&11.000	&P	&1	&2.03	&115.5	&7.4	&	&	&	&20.4	\\
	&	&	&	&	&P	&2	&5.27	&133.0	&8.1	&	&	&	&21.2	\\
\hline
K00124	&11086270	&12.935	&11.919	&11.622	&P	&1	&2.4	&41.3	&6.6	&	&	&	&20.2	\\
\hline
K00126	&5897826	&13.109	&11.977	&11.634	&P	&1	&3.03	&134.7	&6.9	&	&	&	&20.6	\\
	&	&	&	&	&P	&2	&3.54	&85.0	&7.1	&	&	&	&20.8	\\
	&	&	&	&	&P	&3	&4.08	&12.3	&7.8	&	&	&	&21.6	\\
\hline
K00137	&8644288	&13.549	&12.189	&11.756	&P	&1	&4.8	&19.5	&7.9	&	&	&	&21.9	\\
	&	&	&	&	&P	&2	&4.98	&136.3	&7.5	&	&	&	&21.5	\\
	&	&	&	&	&P	&3	&5.44	&10.1	&4.1	&	&	&	&17.8	\\
	&	&	&	&	&P	&4	&5.72	&73.4	&8.2	&	&	&	&22.2	\\
\hline
K00141	&12105051	&13.687	&12.489	&11.986	&A	&1	&1.06	&13.9	&1.2	&1.06	&13.5	&1.4	&14.8	\\
\hline
K00148	&5735762	&13.040	&11.701	&11.221	&P	&1	&2.44	&114.5	&4.9	&	&	&	&18.1	\\
	&	&	&	&	&P	&2	&4.32	&139.7	&3.3	&	&	&	&16.4	\\
	&	&	&	&	&P	&3	&4.39	&107.7	&7.3	&	&	&	&20.7	\\
	&	&	&	&	&P	&4	&5.89	&121.3	&7.0	&	&	&	&20.4	\\
\hline
K00153	&12252424	&13.461	&11.885	&11.255	&P	&1	&5.14	&84.8	&8.3	&5.16	&84.3	&8.1	&22.3	\\
\hline
K00249	&9390653	&14.486	&12.000	&11.154	&P	&1	&4.19	&27.2	&0.7	&4.19	&27.2	&0.7	&14.9	\\
\hline
K00251	&10489206	&14.752	&12.483	&11.682	&P	&1	&3.45	&121.2	&3.9	&3.45	&121.2	&4.1	&17.9	\\
	&	&	&	&	&P	&2	&4.76	&13.3	&6.5	&4.75	&13.4	&6.4	&21.4	\\
\enddata     
\label{table:compstars}
\end{deluxetable*}

\setcounter{table}{3}
\begin{deluxetable*}{crrrrcc  |  rrr | rrr   |  c}
\tablewidth{0pt}
\tablecaption{Stars within 6\arcsec\ of \emph{Kepler} planetary candidates}
\tablehead{
	&		&	&2MASS	&2MASS	&	&  & \multicolumn{3}{c}{J} 	& \multicolumn{3}{c}{Ks}  & Est Kep\\
KOI	&KeplerID	 &Kep	& J	& K	&Inst.\tablenotemark{a}	& Star\tablenotemark{b} &Dist(\arcsec)	&PA($^{\circ}$)\tablenotemark{c}	&$\Delta$ Mag\tablenotemark{d} &Dist(\arcsec)	&PA($^{\circ}$)\tablenotemark{c}	 &$\Delta$ Mag\tablenotemark{d} &  Mag\tablenotemark{e}
		}
\startdata
K00258	&11231334	&9.887	&8.946	&8.682	&A	&1	&0.98	&72.0	&2.5	&1.0	&73.3	&2.5	&12.4	\\
	&	&	&	&	&A	&2	&1.37	&73.2	&3.1	&1.43	&74.6	&2.9	&13.3	\\
\hline
K00261	&5383248	&10.297	&9.259	&8.868	&A	&1	&5.42	&65.2	&7.1	&5.41	&69.7	&6.8	&18.1	\\
\hline
K00263	&10514430	&10.821	&9.429	&9.007	&A	&1	&3.2	&91.2	&0.6	&3.2	&79.0	&0.6	&11.2	\\
	&	&	&	&	&P	&1	&3.19	&91.7	&0.8	&3.19	&91.7	&0.8	&11.4	\\
\hline
K00264	&3097346	&11.551	&10.370	&10.020	&A	&1	&0.45	&36.0	&3.5	&0.48	&41.7	&3.5	&15.0	\\
\hline
K00266	&7375348	&11.472	&10.674	&10.379	&A	&1	&3.62	&35.6	&6.6	&3.62	&22.9	&6.1	&19.3	\\
\hline
K00268	&3425851	&10.560	&9.948	&9.395	&A	&1	&1.76	&87.0	&3.1	&1.72	&87.4	&2.5	&14.9	\\
	&	&	&	&	&A	&2	&2.52	&48.2	&4.5	&2.46	&45.1	&3.9	&16.3	\\
\hline
K00269	&7670943	&10.927	&9.968	&9.753	&A	&1	&	&	&	&2.61	&74.4	&7.6	&20.4	\\
K00270	&6528464	&11.411	&10.088	&9.701	&A	&1	&0.05	&71.3	&-0.0	&0.11	&65.5	&0.1	&11.1	\\
\hline
K00273	&3102384	&11.457	&10.356	&9.967	&A	&1	&5.51	&17.5	&5.8	&5.5	&16.5	&5.3	&18.5	\\
\hline
K00283	&5695396	&11.525	&10.417	&10.079	&P	&1	&5.96	&88.6	&7.9	&5.94	&88.5	&7.8	&19.6	\\
\hline
K00284	&6021275	&11.818	&10.797	&10.423	&P	&1	&0.84	&96.7	&0.3	&0.84	&96.8	&0.3	&12.3	\\
\hline
K00285	&6196457	&11.565	&10.747	&10.403	&P	&1	&1.44	&138.3	&4.2	&1.45	&137.9	&4.1	&16.2	\\
	&	&	&	&	&P	&2	&	&	&	&2.29	&26.8	&6.7	&	\\
\hline
K00292	&11075737	&12.872	&11.743	&11.345	&P	&1	&0.36	&121.8	&2.7	&0.37	&121.8	&2.8	&15.5	\\
\hline
K00303	&5966322	&12.193	&11.019	&10.631	&P	&1	&5.79	&93.8	&7.2	&5.79	&93.7	&7.1	&19.5	\\
\hline
K00306	&6071903	&12.630	&11.257	&10.760	&P	&1	&2.04	&114.4	&2.3	&2.04	&114.5	&2.0	&15.6	\\
	&	&	&	&	&P	&2	&4.52	&32.0	&7.7	&4.53	&32.1	&7.4	&21.0	\\
	&	&	&	&	&P	&3	&5.33	&139.0	&7.0	&5.33	&139.0	&6.4	&20.6	\\
\hline
K00316	&8008067	&12.701	&11.529	&11.166	&P	&1	&5.04	&7.9	&7.1	&5.02	&8.0	&6.7	&20.6	\\
\hline
K00364	&7296438	&10.087	&8.989	&8.644	&P	&1	&5.73	&125.0	&7.7	&5.73	&125.1	&6.9	&18.6	\\
	&	&	&	&	&P	&2	&5.96	&131.7	&7.2	&5.96	&131.7	&6.6	&18.7	\\
\hline
K00372	&6471021	&12.391	&11.294	&10.914	&A	&1	&	&	&	&2.49	&157.8	&8.6	&23.2	\\
	&	&	&	&	&A	&2	&	&	&	&3.56	&56.9	&8.0	&22.4	\\
	&	&	&	&	&A	&3	&	&	&	&4.99	&170.7	&8.2	&22.7	\\
	&	&	&	&	&A	&4	&	&	&	&5.94	&32.7	&4.0	&17.1	\\
\hline
K00377	&3323887	&13.803	&12.710	&12.336	&P	&1	&2.79	&37.9	&6.8	&2.79	&37.8	&6.6	&20.9	\\
	&	&	&	&	&P	&2	&5.9	&91.7	&4.5	&5.89	&91.7	&4.2	&18.9	\\
\hline
K00975	&3632418	&8.224	&7.229	&6.945	&A	&1	&	&	&	&0.72	&132.3	&3.6	&12.9	
\enddata     
\tablenotetext{a} {A=ARIES, P=PHARO.}
\tablenotetext{b} {Each unique companion star detected is numbered. The same number is used for both ARIES and PHARO observations of the same star.}
\tablenotetext{c} {Angle from north. Note that the ARIES angles were determined from the north-east directions of the dither pattern, which sometimes had random jitter, and may differ from the true angle by a few degrees.}
\tablenotetext{d} {Error on the delta magnitude is about 0.01 mag.}
\tablenotetext{e} {$Kp$ magnitude estimated for a dwarf companion using Equation~\ref{eqnBoth} if both $J$ and $Ks$ are available, and otherwise from Equation~\ref{eqnJ} or ~\ref{eqnKs}.}
\tablenotetext{f} {Due to the large, smeared PSF, the parameters for K00013 in ARIES-$J$ are not considered reliable.}
\tablenotetext{g} {Companion distance was near or below the FWHM, so magnitudes were estimated using PSF fitting.}
\label{table:compstars}
\end{deluxetable*}

\clearpage

\clearpage

\end{document}